# Comparison of jump frequencies of $^{111}$In/Cd tracer atoms in Sn$_3$R and In$_3$R phases having the L1$_2$ structure (R = rare-earth)

Megan Lockwood Harberts [a], Benjamin Norman [b], Randal Newhouse [c] and Gary S. Collins [d]

Department of Physics and Astronomy, Washington State University,

Pullman, WA 99164, USA

[a] harberts.2@mps.ohio-state.edu, [b] bmnorman@umich.edu, [c] randynewhouse@gmail.com, [d] collins@wsu.edu

**Abstract.** Measurements were made of jump frequencies of $^{111}$In/Cd tracer atoms on the Sn-sublattice in rare-earth tri-stannides having the L1$_2$ crystal structure via perturbed angular correlation spectroscopy (PAC). Phases studied were Sn$_3$R (R= La, Ce, Pr, Nd, Sm and Gd). Earlier measurements on isostructural rare-earth tri-indides showed that the dominant diffusion mechanism changed along that series [4]. The dominant mechanism was determined by comparing jump frequencies measured at opposing phase boundary compositions (that is, more In-rich and more In-poor). Jump frequencies were observed to be greater at the In-rich boundary composition in light lanthanide indides and greater at the In-poor boundary composition in heavy-lanthanide indides. These observations were attributed to predominance of diffusion via rare-earth vacancies in the former case and indium vacancies in the latter. Contrary to results for the indides, jump frequencies found in the present work are greater for the Sn-poor boundary compositions of the stannides, signaling that diffusive jumps are controlled by Sn-vacancies. Possible origins of these differences in diffusion mechanisms are discussed.

**Introduction**

There is considerable interest in diffusion mechanisms in compounds having increasingly complex structures [1]. Most commonly, diffusion is mediated by atomic vacancies when structures are close-packed and atomic sizes of the different elements are similar, as for the phases studied here. In highly-ordered compounds, the energy cost to make a point defect (antisite atom or vacancy) is large and atom movement can occur only in correlated jump sequences that eliminate lattice disorder produced by jumps. In an A$_3$B compound such as one having the L1$_2$, or Cu$_3$Au, crystal structure, it makes a difference whether structural or thermally activated vacancies are on the A- or B-sublattices. This is illustrated in Fig. 1, showing two hypothetical diffusion mechanisms. At left is shown an A-vacancy jumping on the A-sublattice, without creating lattice defects. However, a B-vacancy must create lattice disorder when jumping to near-neighbor sites. The sequence of six jumps shown eliminates disorder created during the first three jumps by the second three jumps.

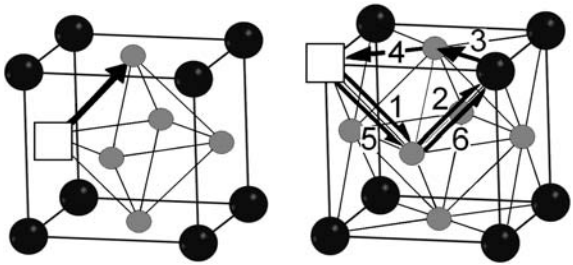

**Fig. 1.** Diffusion mechanisms in an L1$_2$ crystal structure of composition A$_3$B starting from an A-vacancy (left) or B-vacancy (right). (After Ref. 4.)

Microscopic methods can shed considerable light on diffusion mechanisms. Several years ago, some of us showed that the method of perturbed angular correlation of gamma rays (PAC) can be used to measure diffusional jump frequencies of PAC probe atoms in compounds [2]. Measurements were made using the well-known $^{111}$In/Cd probe. For further details about PAC spectroscopy, see Ref. 3. Atomic jumps that lead to reorientation or change in magnitude of electric field gradients at probe nuclei cause relaxation of the nuclear quadrupole interaction. The relaxation is observed for the PAC level of the daughter $^{111}$Cd nucleus, so that measured jump frequencies are of Cd-tracer atoms as *dilute impurities* in the stannides and indides. Qualitatively, the relaxation appears as damping of the quadrupole perturbation function, and can be fitted using stochastic jump models to obtain accurate values of the mean jump frequency (inverse of the mean residence time). This was illustrated by measurements on a pair of samples of In$_3$La. While In$_3$La appears as a *line compound* in binary phase diagrams, suggesting that it has a fixed 75:25 stoichiometry, all compounds at non-zero temperature have a phase field of finite width, no matter how small. The two samples of In$_3$La were made to have compositions at the opposing phase boundaries, In-richer and In-poorer. It was found that jump frequencies for the In-richer phase boundary composition were greater by factors of 10-100, depending on temperature. The significance of this difference was not fully appreciated at the time.

Measurements since were extended to the entire series of In$_3$R phases (R= rare earth) [4], from which it was found that jump frequencies were greater at the In-rich boundary for light lanthanide indides (La, Ce, Pr, Nd) but greater at the In-poor boundary for heavy lanthanides (Sm to Lu). This difference was shown to signal a change in the vacancy that is primarily responsible for diffusion of probe atoms on the In-sublattice. It can be shown that vacancy concentrations, whether structural or thermal, change monotonically with compositions. Thus, the In-vacancy concentration increases as the In-composition decreases and *vice versa*. Consequently, the measurements showed that In-vacancies are the principal driving force for diffusion in the heavy lanthanide indides, and R-vacancies in the light lanthanide indides.

As to identifying the precise diffusion mechanisms, the A-sublattice diffusion mechanism shown in Fig. 1, left, is almost certainly the dominant A-vacancy mechanism since diffusion then occurs without creation of additional point defects such as antisite atoms. The B-vacancy, six-jump cycle mechanism is the most likely other mechanism. These and other hypothetical mechanisms have been discussed in [1,2,4,5]. The change in mechanism between light and heavy lanthanide indides is probably caused by gradual changes in partial formation enthalpies of the four elementary point defects: A- and B-vacancies and A$_B$ and B$_A$ antisite atoms. It is noteworthy that the changeover reported in Ref. 4 appears to be correlated with lattice parameter. The lattice parameter decreases significantly along the lanthanide series due to lanthanide contraction, as shown in Fig. 3 of Ref. 4 and below in Fig. 7. Calculations are in progress in this laboratory using the WIEN2K code [6] to test this hypothesis.

It is of interest to see if there are similar changes in diffusion mechanism in other series of L1$_2$ phases with rare-earth elements. Possible series include Al$_3$R, Ga$_3$R and Sn$_3$R. We attempted measurements at opposing phase boundary compositions in most phases having the L1$_2$ structure, including aluminides of Er, Tm, Lu and Yb [7], and gallides of Dy, Er and Lu [8]. However, for R-poor compositions in all seven phases, it was found that the $^{111}$In/Cd probe atoms transferred to the R-sites, which are cubic, making it impossible to measure jump frequencies on the aluminum or gallium sublattice at both boundary compositions.

In this paper, we present results of measurements on six Sn$_3$R phases, of which measurements were made at opposing phase boundary compositions for four: LaSn$_3$, CeSn$_3$, SmSn$_3$ and GdSn$_3$. An earlier account of measurements on LaSn$_3$ is in Ref. 9.

**Experiments and Results**

High purity metals were melted together with trace amounts of [111]In activity under argon to make ~100 mg spherical samples. Pairs of samples were made having off-stoichiometric Sn-rich and Sn-poor compositions, with typically about 23 and 27 at.% rare-earth, making the samples two-phased with the L1$_2$ phases at their boundary compositions. Small signals from neighboring phases were fitted and did not disturb the analysis. Measurements were made using four-counter PAC spectrometers of standard design [3,2,9]. Representative spectra are shown in Fig. 2.

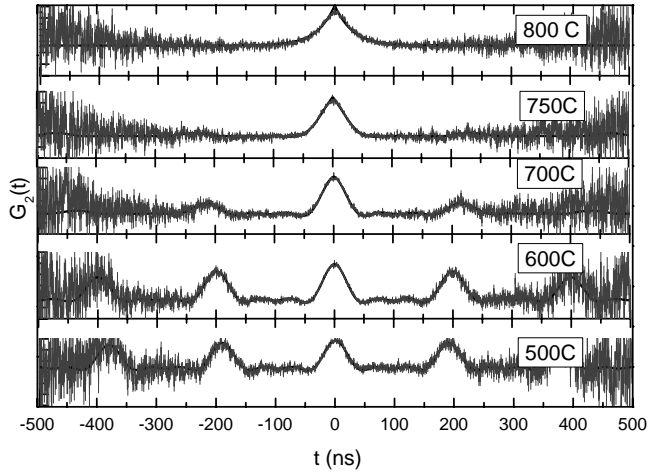

**Fig. 2.** PAC perturbation functions for PAC probes on the Sn-sublattice in Sn-poor SmSn$_3$ measured at the indicated temperatures. Increased damping at higher temperatures is attributed to diffusional relaxation caused by reorientations of electric field gradient tensors in each jump. (Measurements at negative times are independent data mirroring measurements at positive time.)

Spectra were fitted with a quadrupole perturbation function $G_2(t)$ of the form

$$G_2(t) = \exp(-wt) G_2^{static}(t), \tag{1}$$

which is a good approximation to the exact dynamically damped perturbation function in the slow fluctuation regime, that is, when the jump frequency is less than the quadrupole interaction frequency [2, 10]. In the present situation, with reorientations of the EFG tensor by 90 degrees in each jump, the factor $w$ is simply the mean jump frequency (inverse of the mean residence time of the probe atom on a site of the A-sublattice). Arrhenius plots of jump frequencies measured for six Sn-poor stannides are shown in Fig. 3.

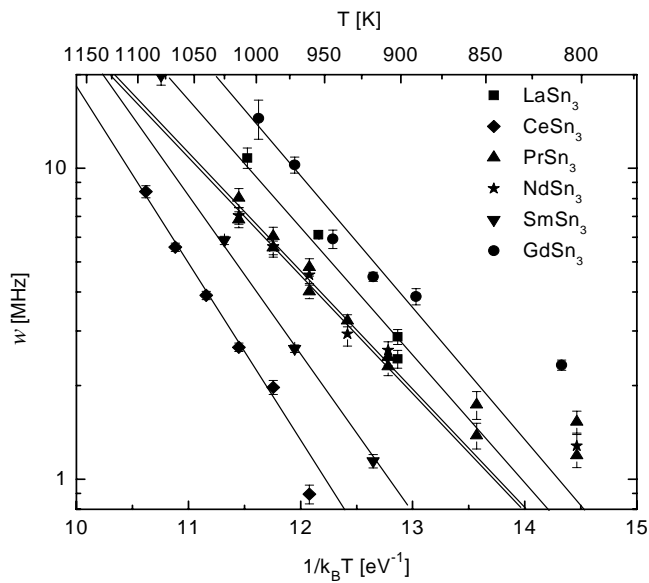

**Fig. 3.** Arrhenius plots of jump frequencies of [111]In/Cd probe atoms on Sn-sublattices in the indicated stannide phases, with compositions at the Sn-poor phase boundaries.

The activation enthalpies are all similar, although jump frequencies differ by up to a factor of ten between the phases. Fig. 4 shows similar Arrhenius plots for Sn-poor and Sn-rich samples of LaSn$_3$ and CeSn$_3$ and Fig. 5 shows results for SmSn$_3$ and GdSn$_3$. Fig. 6 compares jump frequencies in lanthanum stannide and indide.

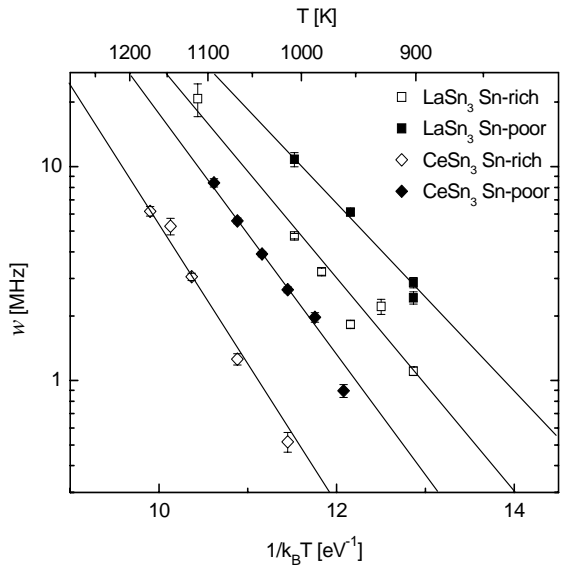

**Fig. 4.** Arrhenius plots of jump frequencies of $^{111}$In/Cd probe atoms on the Sn-sublattices in Sn-rich and Sn-poor samples of LaSn$_3$ and CeSn$_3$. It can be seen that jump frequencies are greater for the Sn-poor boundary compositions.

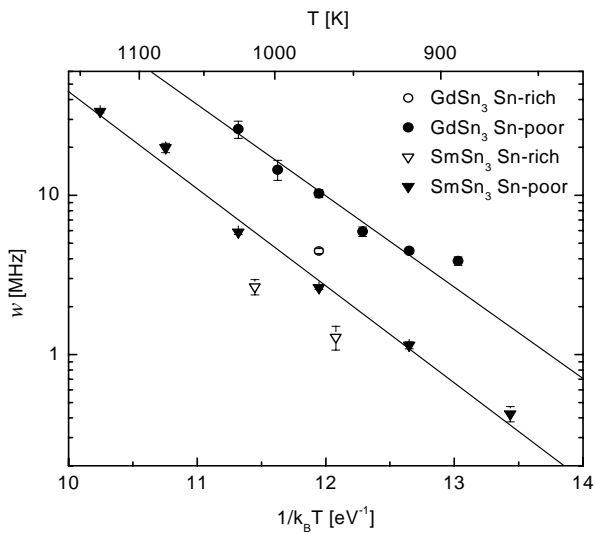

**Fig. 5.** Arrhenius plots of jump frequencies of $^{111}$In/Cd probe atoms on the Sn-sublattices in Sn-rich and Sn-poor samples of SmSn$_3$ and GdSn$_3$. It can be seen that jump frequencies are greater for Sn-poor boundary compositions. Data for Sn-rich boundary compositions consist of only a few points due to measurement difficulties, but clearly show that jump frequencies are greater at the Sn-poor boundary compositions by a factor of 2-3.

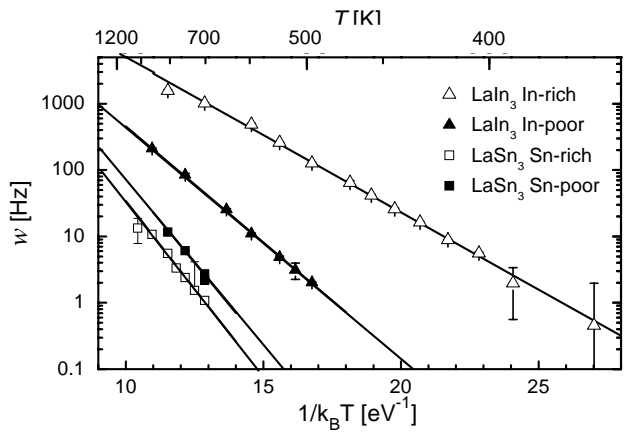

**Fig. 6.** Comparison of jump frequencies in lanthanum stannides and indides.

It can be seen that jump frequencies are greater for the In-rich boundary composition in In$_3$La, but greater for the Sn-poor composition of Sn$_3$La. In addition, it is noteworthy that the jump frequencies in the light lanthanide indides differ by factors as large as 10-100, but only by a factor of 3 in the stannides and heavy lanthanide indides. Table 1 lists fitted jump-frequency activation enthalpies and prefactors obtained from fits of the individual sets of data using

$$w = w_0 \exp(-Q/k_B T),  \qquad (2)$$

in which $Q$ is the jump-frequency activation enthalpy and $w_0$ is the jump-frequency prefactor.

Table 1. Jump-frequency activation enthalpies and prefactors

| Phase | $Q$ [eV] | $w_0$ [THz] |
|---|---|---|
| LaSn$_3$ Sn-Rich | 1.29(5) | $14 \pm^{12}_{7}$ |
| LaSn$_3$ Sn-Poor | 1.09(10) | $3.4 \pm^{8.2}_{2.4}$ |
| CeSn$_3$ Sn-Rich | 1.63(7) | $69 \pm^{76}_{36}$ |
| CeSn$_3$ Sn-Poor | 1.41(8) | $26 \pm^{42}_{16}$ |
| PrSn$_3$ Sn-Poor | 0.84(5) | $0.11 \pm^{0.09}_{0.05}$ |
| NdSn$_3$ Sn-Poor | 0.78(7) | $0.054 \pm^{0.066}_{0.030}$ |
| SmSn$_3$ Sn-Poor | 1.49(7) | $130 \pm^{150}_{70}$ |
| GdSn$_3$ Sn-Poor | 1.05(15) | $3 \pm^{15}_{2}$ |

All activation enthalpies are in the range 0.8-1.6 eV, and prefactors in the range 0.1-130 THz. The prefactors appear consistent with a vacancy diffusion mechanism since the magnitudes are of the order of a jump attempt frequency that is equal to the vibrational frequency of an atom in a solid.

**Discussion**

**Difference in diffusion mechanism in stannides and indides with light lanthanide elements.** A convenient factor that can be used to represent results of an Arrhenius fit such as in Figs. 3-6 is the temperature $T_{10}$ at with the jump frequency equals 10 MHz [4]. A jump frequency of 10 MHz corresponds to a diffusivity of about $10^{-13}$ m$^2$/sec. Fig. 7 shows such a plot of the inverse temperatures $1/k_B T_{10}$ for aluminides, gallides, indides and stannides that have been studied. Several points are noteworthy:

1. As can be seen, jump frequencies are 10 MHz for most phases at roughly 900 K.

2. Measurements were made at both R-poor and R-rich boundary compositions (R= rare earth) for the indides and some stannides. Measurements in aluminides and gallides could only be made for R-rich boundary compositions because of a strong site preference of indium solutes for rare-earth sites in R-poor samples (see Ref. [11]).

3. Behavior of the light lanthanide indides appears anomalous, with 10 MHz frequencies observed at much lower temperatures. There is also a change in diffusion mechanism from one exhibiting greater jump frequencies at the In-poor phase boundary composition for the heavy lanthanide indides (designated B in the plot) to one with greater frequencies at the In-rich boundary composition (A) for the light lanthanide indides.

4. With the exception of the light lanthanide indides, it can be shown that jump frequencies are greater by a factor of approximately 3 at rare-earth-rich phase boundary compositions (cf. Figs. 4-6 and Ref. [4]).

5. Except for the anomalous behavior of the light lanthanide indides, there is no obvious dependence of jump frequencies on the lattice parameter.

6. The Cd-tracer atoms on which all jump frequencies are measured are impurities, so that interactions will exist between tracer atoms and vacancy defects that differ from those of host atoms in the stannides, indide, aluminide and gallide series. Thus, the measured jump frequencies do not necessarily correspond to those of host elements.

Point 3 is consistent with the A-sublattice vacancy diffusion mechanism [5], shown in Fig. 1 (left). The deviation observed for the light-lanthanide indides appears exceptional, in particular since it is not observed in the stannides.

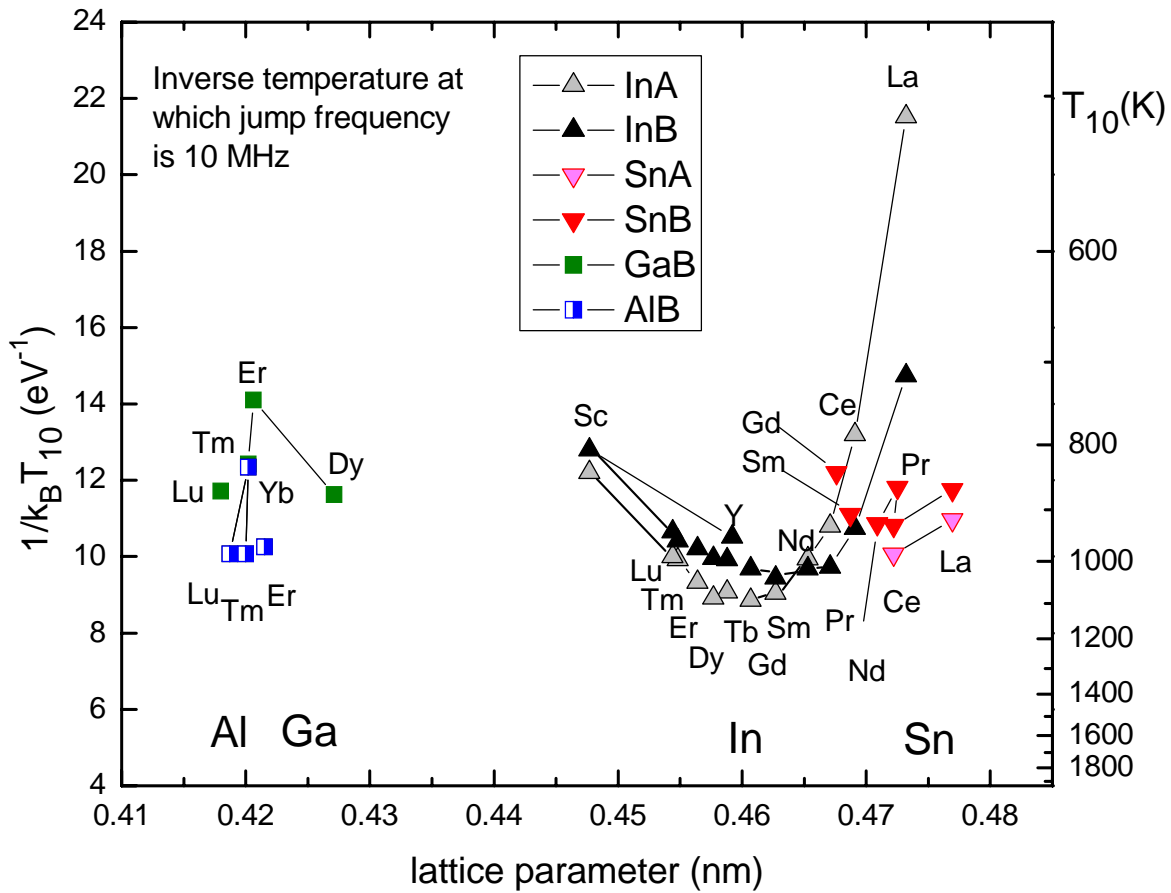

**Fig. 7.** Inverse temperatures at which jump frequencies of $^{111}$In/Cd probe atoms equal 10 MHz, plotted versus lattice parameters of the indicated rare-earth aluminide, gallide, indide and stannides phases having $L1_2$ structure. Lattice parameters increase in going from the aluminides to gallides, indides and stannides. A and B designate, respectively, R-poor and R-rich boundary compositions.

**Boundary compositions and structural point defects**. To our knowledge, there have been no measurements of compositions of opposing phase boundaries for the "line" compounds considered in the present study. Lack of observable inhomogeneous broadening at low temperature in In-rich

In$_3$La suggested that the In-rich boundary was within roughly 0.2 at.% of the stoichiometric composition [2]. One can obtain a qualitative idea of boundary compositions from the binary phase diagrams [12]. For both the indide and stannide phases, melting temperatures of the L1$_2$ phases are roughly 1100 °C, much greater than melting temperatures of the adjacent, terminal phases of indium and tin metals. The more R-rich adjoining phases all have melting temperatures close to those of the L1$_2$ phases.

Consequently, ignoring entropy differences, one can surmise that free energy curves of the three phases (terminal, L1$_2$ and more R-rich), will have tangents in the usual construction [13] that intersect the free energy curve of the L1$_2$ phase in such a way that the R-rich boundary composition will be very close to the stoichiometric composition whereas the R-poor boundary composition will exhibit a deviation from stoichiometry that is much greater. For In$_3$R, for example, this means that there will be structural defects, R-vacancies and/or In$_R$ antisite atoms, at the In-rich boundary composition but few structural defects at the In-poor boundary composition. Of course there will be thermally activated defects that may include vacancies as well.

Accordingly, a possible explanation for the increase in diffusion mediated by R-vacancies in the light lanthanide indides is an anomalous increase in the relative proportion of R-vacancies relative to In$_R$ antisite atoms. But then one has to assume that proportions of In- or Sn-antisite atoms are much greater than R-vacancies in the stannides and heavy lanthanide indides.

**A non-vacancy diffusion mechanism**? Jump frequencies in the light lanthanide indides are greater by factors of 1000 than in the heavy lanthanide indides (see, for example, Ref. [9] and [14]). It is hard to conceive how this can be caused simply by large differences in vacancy concentration and/or large decreases in the activation enthalpy for vacancy migration. Consequently, a non-vacancy diffusion mechanism might be operative, for example the defect-free exchange mechanism of Zener [15], in which rings of three or four indium atoms would rotate as units by ratcheting. However, it is unclear why such a mechanism should not also be active in the stannides nor why it should not operate equally well at both In-rich and In-poor boundary compositions.

**Summary**

PAC measurements were made of frequencies of jumping of In/Cd probe atoms on the Sn-sublattice in a series of lanthanide stannides having the L1$_2$ structure. Jump frequencies were observed to be greater for Sn-poor phase boundary compositions than for Sn-rich ones, in parallel with observations for heavy-lanthanide indides. Having the greater jump frequency at the Sn-poor boundary indicates that diffusion of the tracer atoms is driven by Sn-vacancies. For similar light-lanthanide indides, an enormous enhancement in jump frequencies had been observed when compared with heavy-lanthanide indides and, at the same time, the jump frequency was observed to be greater at the In-rich phase boundary. Having the greater jump frequency at the In-rich boundary signals that diffusion of tracer atoms on the In-sublattice is driven by rare-earth vacancies. Possible explanations for the differences in behavior include changing proportions of structural (or thermal) defects along the series of light-lanthanide indides, favoring R-vacancies, or even a non-vacancy diffusion mechanism.

**Acknowledgments**

This work was supported in part by the National Science Foundation under grants DMR 05-04843 and 09-04096 (Metals Program). We thank Matthew O. Zacate for discussions and comments.